\documentclass[prb,onecolumn,11pt]{revtex4}
\usepackage{amsmath}
\usepackage{graphicx}
\usepackage{dcolumn}
\usepackage{bm}

\begin{document}

\title{Magnetization Plateaux in the Antiferromagnetic Ising Chain with
Single-Ion Anisotropy\thanks{%
We would like to thank CNPq (Brazilian agency) for financial support.}}
\author{Fabian Litaiff$^{a}$, J. Ricardo de Sousa$^{a,b}$ and N. S. Branco$%
^{c}$}
\affiliation{$^{a}$Departamento de F\'{\i}sica, Universidade Federal do Amazonas, 3000
Japiim, 69077-000, Manaus-AM, Brazil\\
$^{b}$ Departamento de F\'{\i}sica, ICEx, Universidade Federal de Minas
Gerais, Av. Ant\^{o}nio Carlos 6627, CP 702, 30123-970, Belo Horizonte-MG,
Brazil.\\
$^{c}$ Departamento de F\'{\i}sica, Universidade Federal de Santa Catarina,
88040-900, Florian\'{o}polis-SC, Brazil}

\begin{abstract}
Two one-dimensional spin-$1$ antiferromagnetic Ising models with a
single-ion anisotropy under external magnetic field at low temperatures are
exactly investigated by the transfer-matrix technique. The
magnetization per spin ($m$) is obtained for the two types of models
(denoted by model 1 and 2) as an explicit function of the magnetic field ($H$%
) and of the anisotropy parameter ($D$). Model $1$ is an extension of
the recently one treated by Ohanyan and Ananikian [\emph{Phys. Lett. A} \textbf{%
307} (2003) 76]: we have generalized their model to the spin-$1$ case
and a single-ion anisotropy term have been included. In the limit of positive (or null) anisotropy ($D\geq 0$) and
strong antiferromagnetic coupling ($\alpha =J_{A}/J_{F}\geq 3$) the $m \times H$ curves
are qualitatively the same as for the
spin $S=1/2$ case, with the presence of only one plateau at $m/m_{sat}=1/3$. On
the other hand, for negative anisotropy ($D<0$) we observe more
plateaux ($m=1/6$ and $2/3$), which depend on the values of $D$ and $%
\alpha $. The second model (model $2$) is the same as the one recently studied
by Chen et al. [\emph{J. Mag. Mag. Mat. }\textbf{262} (2003) 258)] using
Monte Carlo simulation; here, the model is treated within an exact
transfer-matrix framework.

PACS NUMBER: 75.10.Hk, 75.10.FPg, 75.40.Cx, 75.50.Ee.

\end{abstract}

\maketitle

\section{Introduction}

For some one-dimensional (1D) antiferromagnets at low temperatures, it has been
observed a spin gap, which is induced by a finite magnetic field,
and a plateaux structure appears in the magnetization process.
Experimentally, the magnetization plateaux were observed in high-field
measurements of several magnetic materials such as the quasi one-dimensional
compounds SrCu$_{2}$O$_{3}$\cite{1}, Y$_{2}$BaNiO$_{5}$\cite{2}, Ni(C$_{2}$H$%
_{8}$N$_{2}$)$_{2}$NO$_{2}$ClO$_{4}$ (abbreviated NENP)\cite{3,4}, and
Cu(NO$_{3}$)$_{2}2.5$H$_{2}$O\cite{5}, the triangular antiferromagnets C$%
_{6} $Eu\cite{6}, CsCuCl$_{3}$\cite{7} and RbFe(MoO$_{4}$)$_{2}$\cite{8}, and
the quasi two-dimensional compound, with a  Shastry-Sutherland
lattice structure, SrCu$_{2}$(BO$_{3}$)$_{2}$\cite{9}. The mechanism 
for the appearance of these
magnetization plateaux in quasi one-dimensional spin chains are \textit{%
dimerization}, \textit{frustration}, \textit{single-ion anisotropy}, \textit{%
periodic field} and so on.

From a general view point, Oshikawa et al.\cite{10} concluded that the
necessary condition for the magnetization plateaux in spin-$S$ chains is
$Q(S-m)=$integer, where $Q$ is the spatial periodicity of the
magnetic ground state and $m$ is the magnetization per site. For some
range of the magnetic field $H$ (i.e., $H_{1}<H<H_{2}$), the system ceases
responding to its increase and a plateau is formed in the magnetization versus
the magnetic field curve. The values of $m$ at which the plateaux
appears are sensitive to small changes in the parameters of the model
and are not only restricted to integer spin (Haldane conjecture)\cite{11}. 

In the $S=1/2$ antiferromagnetic Heisenberg model on a triangular lattice, a
magnetization plateau was found at $m/m_{sat}=1/3$\cite{6,7,8,12}. In an $%
S=1/2$ trimerized Heisenberg model\cite{13}, the plateau appears at $%
m/m_{sat}=1/6$. Recently, plateaux at $m/m_{sat}=1/8$ and $1/4$ have been
observed in the SrCu$_{2}$(BO$_{3}$)$_{2}$\cite{9}, which has a
Shastry-Sutherland lattice structure. However, \textit{irrational} values have not been found, at
least so far. Theoretically, various other models with spin $S=1/2$ have been
proposed to describe the magnetization plateaux. One of the first models was
introduced by Hida\cite{14}, where a Heisenberg chain was
considered, with antiferromagnetically coupled ferromagnetic trimers ($p=3$).
The three-dimerized Hamiltonian proposed by Hida to
describe the $3$CuCl$_{2}$.$2$ dioxane compound is given
by 
\begin{equation}
\mathcal{H}=\mathcal{H}^{trim}+\mathcal{H}^{int}+\mathcal{H}^{Zeeman},
\label{1}
\end{equation}%
with 
\begin{equation}
\mathcal{H}^{trim}=-J_{F}\sum\limits_{i}\left( \mathbf{S}_{i}\cdot \mathbf{%
\tau }_{i}+\mathbf{\tau }_{i}\cdot \mathbf{\sigma }_{i}\right) ,  \label{2}
\end{equation}%
\begin{equation}
\mathcal{H}^{int}=J_{A}\sum\limits_{i}\mathbf{\sigma }_{i}\cdot \mathbf{S}%
_{i+1},  \label{3}
\end{equation}%
and 
\begin{equation}
\mathcal{H}^{Zeeman}=-\mu _{B}H\sum_{i}\left( S_{i}^{z}+\sigma _{i}^{z}+\tau
_{i}^{z}\right) ,  \label{4}
\end{equation}%
where $J_{A}$ and $J_{F}$ are the antiferromagnetic and ferromagnetic
interactions, respectively, $\mathbf{S}_{i},\mathbf{\tau }_{i}$ and $\mathbf{%
\sigma }_{i}$ are the $S=1/2$ spin operators at site $i$, $\mu _{B}$ is the
Bohr magneton and $H$ is the magnetic field. Using exact diagonalization of
finite systems, Hida obtained, for $J_{F}$ comparable to or smaller than $%
J_{A} $, a plateau at $m/m_{sat}=1/3$. The plateau mechanism was
considered to be a purely quantum phenomenon, where the concepts of magnetic
quasiparticles and strong quantum fluctuations are regarded to be of
major importance for understanding the process. On the other hand,
Ohanyan and Ananikian\cite{15} have recently studied the Hida model by using the
transfer-matrix technique, replacing the spin operators (%
$\mathbf{S}_{i},\mathbf{\tau }_{i}$ and $\mathbf{\sigma }_{i}$) by Ising
variables ($S_{i}^{z},\sigma _{i}^{z},\tau _{i}^{z}$). It was shown that, for this
classical model and for $T=0$ (ground state) and $J_{A}\geq 3J_{F}$ (strong
antiferromagnetic coupling), a magnetization curve with plateau at $%
m/m_{sat}=1/3$ is observed, indicating that the appearance of plateaux is not a quantum
manifestation, but may be caused by the stability of spatially modulated spin
structures.

Another model which presents magnetization plateaux is the one-dimensional spin-$1$
antiferromagnetic Heisenberg with single-ion anisotropy\cite{16}. This model
is described by the following Hamiltonian 
\begin{equation}
\mathcal{H}=J\sum\limits_{i}\mathbf{S}_{i}\cdot \mathbf{S}_{i+1}-\mu
_{B}H\sum_{i}S_{i}^{z}+D\sum_{i}\left( S_{i}^{z}\right) ^{2},  \label{5}
\end{equation}%
where $D$ is the single-ion anisotropy. For $D=0$, the
ground state is a singlet and the lowest excitation is a triplet
(Haldane conjecture\cite{11}); increasing $D$, the triplet splits into a
higher-energy singlet and a lower-lying doublet, with the Haldane gap for
$D=0$, $\Delta (0)$, spliting into two gaps, as observed in neutron
scattering of NENP\cite{17}. The Haldane gap for general $D$, $\Delta (D)$, presents 
two different behaviors: for $D>D_{c}=J$, it increases with $D$, while
for $D<D_{c}$ $\Delta (D)$ decreases as $D$ increases.

Recently, spin $S\geq 1$ Ising antiferromagnetic chains with single-ion
anisotropy have been studied by using classical Monte Carlo simulation\cite%
{18} and it was observed the presence of $2S+1$ plateaux for $D>0$. Essentially, these
classical models are obtained replacing the spin operators ($\mathbf{S}_{i}$) 
by Ising variables ($S_{i}^{z}$) in
Hamiltonian (5). From a theoretical point view, the model studied by Chen, et.
al.\cite{18} represent the 1D antiferromagnetic Blume-Capel model\cite{19},
and it was observed also two different critical behaviors, which depend on the
anisotropy parameter $D$ ($D<D_{c}$ and $D>D_{c}$, where $D_{c}=J$).

The purpose of this work is to obtain exact results for two classical models with
spin $S=1$ and in the presence of a single-ion anisotropy. In Section 2 the 1D models are
presented and exactly solved by the transfer-matrix technique. The
magnetization plateaux and ground-state phase diagrams are discussed in
Section 3. Finally, the last section is devoted to conclusions.

\section{Models and Formalism}

The transfer-matrix technique was proposed years ago by Kramers and Wannier%
\cite{20,21}, and it formed the basis for Onsager's solution\cite{22} of the
two-dimensional Ising model. In this section, we use this technique to
obtain exact results for two one-dimensional
models, in order to analyze the magnetization plateau mechanism.

\subsection{Model 1: Three-dimerized chain}

The first model we study is described by the following
Hamiltonian:
\begin{eqnarray}
\mathcal{H}_1 &=&-J_F\sum\limits_i\left( S_i^z\cdot \tau _i^z+\sigma
_i^z\cdot \tau _i^z-\alpha \sigma _i^z\cdot S_{i+1}^z\right) -\mu
_BH\sum_i\left( S_i^z+\tau _i^z+\sigma _i^z\right) -  \notag \\
&&D\sum_i\left[ \left( S_i^z\right) ^2+\left( \tau _i^z\right) ^2+\left(
\sigma _i^z\right) ^2\right] ,  \label{6}
\end{eqnarray}
where $\alpha =J_A/J_F$ and the spin variables $S_i^z,\tau _i^z$ and $%
\sigma _i^z$ can assume the values $-1,0,1$. The above Hamiltonian represents a
nonuniform spin system in which ferromagnetic trimers composed of $S=1$
spins ($S_i^z,\tau _i^z$ and $\sigma _i^z$) are coupled
antiferromagnetically in one dimension, in the presence of a magnetic field ($%
H$) and single-ion anisotropy ($D$). In the limit $\alpha \rightarrow 0$
(strong intratrimer ferromagnetic interaction), the variables $S_i^z,\tau
_i^z$ and $\sigma _i^z$ form a single spin $\xi _i$ with magnitude $3$.
Thus, the system can be approximated by a spin $S=3$ antiferromagnetic
Blume-Capel chain.

The transfer-matrix technique is based in the calculations of the
eigenvalues $\{\lambda _{i}\}$, determined from the solution of the
secular equation 
\begin{equation}
\det (W_{1}-\lambda I)=0,  \label{7}
\end{equation}%
where $I$ is the identity matrix $3\times 3$ and $W_{1}$ the Wannier matrix,
with the elements defined by 
\begin{equation}
W_{1}(S,S^{\prime })=\sum\limits_{\sigma ,\tau =0,\pm 1}\exp [a(\tau
)S+dS^{2}+b(\sigma )S^{\prime }+c(\tau ,\sigma )],  \label{8}
\end{equation}%
with 
\begin{equation}
a(\tau )=\beta J_{F}\tau +\beta \mu _{B}H,  \label{9}
\end{equation}%
\begin{equation}
b(\sigma )=-\alpha \beta J_{F}\sigma ,  \label{10}
\end{equation}%
\begin{equation}
c(\tau ,\sigma )=\beta J_{F}\sigma \tau +\beta \mu _{B}H(\tau +\sigma
)+\beta D(\tau ^{2}+\sigma ^{2}),  \label{11}
\end{equation}%
and 
\begin{equation}
d=\beta D,  \label{12}
\end{equation}%
where $S,S^{\prime }=0,\pm 1$.

Using properties of the matrix trace, the partition function $Z=Tr(W^{N})$
can be written as a sum of the $N$th power of the eigenvalues $\{\lambda
_{i}\}$ obtained from Eq.(7), i.e., 
\begin{equation}
Z=\sum\limits_{i=1}^{3}\lambda _{i}^{N}.  \label{13}
\end{equation}

In the thermodynamic limit ($N\rightarrow \infty $), the free energy,
magnetization, magnetic susceptibility and specific heat (per atom) are
expressed in terms of maximum eigenvalue $\lambda _{\max }$, respectively, as
\begin{equation}
f=\frac{-T}{3}\ln \lambda _{\max },
\label{14}
\end{equation}%
\begin{equation}
m=\frac{T}{3\lambda _{\max }}\frac{\partial \lambda _{\max }}{\partial H},
\label{15}
\end{equation}%
\begin{equation}
\chi =\frac{\partial m}{\partial H}=\frac{T}{3}\frac{\partial }{\partial H}(%
\frac{1}{\lambda _{\max }}\frac{\partial \lambda _{\max }}{\partial H}),
\label{16}
\end{equation}%
and 
\begin{equation}
c=\frac{2T}{3\lambda _{\max }}\frac{\partial \lambda _{\max }}{\partial T}%
+T^{2}\frac{\partial }{\partial T}(\frac{1}{\lambda _{\max }}\frac{\partial
\lambda _{\max }}{\partial T}),  \label{17}
\end{equation}%
where the factor $1/3$ was introduced because there are three spins in each
site of the chain, and the maximum eigenvalue $\lambda _{\max }$ is given by 
\begin{equation}
\lambda _{\max }=-\frac{A}{3}+2\sqrt{Q}\cos (\frac{\theta }{3}),  \label{18}
\end{equation}%
with 
\begin{equation}
A=W_{1}(1,1)+W_{1}(0,0)+W_{1}(-1,-1)=Tr(W_{1}),  \label{19}
\end{equation}%
\begin{equation}
Q=\frac{A+3B}{9},  \label{20}
\end{equation}%
\begin{eqnarray}
B &=&W_{1}(1,0)W_{1}(0,1)+W_{1}(1,-1)W_{1}(-1,1)+W_{1}(-1,0)W_{1}(0,-1)- 
\notag \\
&&W_{1}(1,1)W_{1}(0,0)-W_{1}(1,1)W_{1}(-1,-1)-W_{1}(0,0)W_{1}(-1,-1),
\label{21}
\end{eqnarray}%
\begin{equation}
R=\frac{9AB-27C-2A^{3}}{54},  \label{22}
\end{equation}%
\begin{equation}
C=-\det (W_{1}),  \label{23}
\end{equation}%
and 
\begin{equation}
\theta =\arccos (\frac{C}{Q^{3/2}}).  \label{24}
\end{equation}%
Replacing the value of $\lambda _{\max }$ given by Eq. (18) in Eqs.(14)-(17) we obtain all the thermodynamic
properties of a classical three-dimerized chain with single-ion anisotropy.

\subsection{Model 2: Antiferromagnetic Blume-Capel chain}

The second model we study is described by the following Hamiltonian: 
\begin{equation}
\mathcal{H}_{2}=J\sum\limits_{i}S_{i}^{z}S_{i+1}^{z}-\mu
_{B}H\sum_{i}S_{i}^{z}+D\sum\limits_{i}(S_{i}^{z})^{2}.  \label{25}
\end{equation}%
which is treated also through the transfer-matrix technique\cite{20,21}.

The Wannier matrix elements in this case are given by 
\begin{equation}
W_{2}(S,S^{\prime })=\exp (-\beta JSS^{\prime }-\beta DS^{2}+\beta \mu
_{B}HS),  \label{26}
\end{equation}%
where the maximum eigenvalue $\lambda _{\max }$ is analogous to Eq.(18)
with $W_{1}(S,S^{\prime })$ replaced by $W_{2}(S,S^{\prime })$ in Eqs.(19),
(21) and (23). 

The model given by Eq. (25) has been recently treated using Monte Carlo
simulation \cite{18}. Using the transfer-matrix technique, which allows for
an exact solution of the model, we can compare the results obtained by both
procedures cited above.

\section{Results and discussion}

At low temperatures, the behavior of the magnetization for model $1$
as a function of the magnetic field depends on the ratio of coupling
constants, $\alpha =J_{A}/J_{F}$, and on the reduced anisotropy parameter, $\delta
=D/J_{F}$.

For $\delta \geq 0$, the
qualitative results are the same as those obtained by obtained Ohanyan and Ananikian\cite{15},
namely: for $\alpha >\alpha _{c}(\delta )$ (strong antiferromagnetic coupling)
and $T=0$ a magnetization plateau appears at $m=1/3$, for magnetic fileds in the interval
$H\in \lbrack H_{c_{1}},H_{c_{2}}]$, where $h_{c_{1}}\equiv
H_{c_{1}}/J_{F}=2.0$ and $h_{c_{2}}\equiv H_{c_{2}}/J_{F}=\alpha -1.0$. We
have obtained, in this strong antiferromagnetic regime and positive
anisotropy, that the value of the critical ratio $\alpha$, $\alpha_c$, does not depend on $\delta $,
i.e., $\alpha _{c}(\delta )=3.0$. For infinity anisotropy ($\delta
\rightarrow \infty $) our results reduce to the case of the three-dimerized
Ising chain with spin $1/2$. So, in this case, we obtain the
same magnetization plateaux as in Ref.14.


The ground state ($T=0$), in the absence of a magnetic field ($H=0$), is the
antiferromagnetic spatially modulated structure, in which trimers of
spins pointing \textit{up} ($S_{i}=1$) alternate with trimers of
spins pointing \textit{down} ($S_{i}=-1$) (i.e., $......\uparrow \uparrow
\uparrow \downarrow \downarrow \downarrow \uparrow \uparrow \uparrow ......$
and so on), for all values of $\delta \geq 0$ and $\alpha $. We denote this
modulated phase by $\left\langle 3\right\rangle $. In the low field region ($%
H<H_{c_{1}}$), no magnetization is observed ($m=0$). When the magnitude of
the external magnetic field increases, at the critical value $H_{c_{1}}$ the
system passes from its ground state $\left\langle 3\right\rangle $ to the
novel spatially modulated structure $\left\langle 3111\right\rangle $, in
which the periodic sequence of spins consists of one trimer pointing along
the field, in the spin state $S_{i}=1$ ($\uparrow \uparrow \uparrow $), and
another trimer with alternating orientation of spins ($\downarrow \uparrow
\downarrow $). In this spin state $\left\langle 3111\right\rangle $,
$m=1/3$ and increases discontinuously to the
saturation value $m=1$ at the second critical field $%
h_{c_{2}}=\alpha -1.0$. In the above one-dimensional model, the
indispensable condition for the appearance of the plateau at $m=1/3$ is the
strong antiferromagnetic coupling, characterized by $\alpha >3.0$. Therefore,
no plateau is found for $\alpha \leq 3.0$.


On the other hand, when the anisotropy is negative the spin state $S_{i}=0$
(represented by $\bigcirc $) is energetically favorable, when compared
to the  spin states $S_{i}=1$ and $-1$. For
certain values of $\alpha $ and $\delta <0$, we can observe various types of
phase transition: first, a flip of the central spin to the state $S_{i}=0$
with a modulated structure $\left\langle 3101\right\rangle $ (corresponding to the
spin configuration $......\uparrow \uparrow \uparrow \downarrow \bigcirc
\downarrow \uparrow \uparrow \uparrow \downarrow \bigcirc \downarrow .....$)
and magnetization $m=1/6$; second, a flip of the central spin to the state $%
S_{i}=1$ with a modulated structure $\left\langle 3111\right\rangle $
(corresponding to the spin configuration $......\uparrow \uparrow \uparrow
\downarrow \uparrow \downarrow \uparrow \uparrow \uparrow \downarrow
\uparrow \downarrow .....$) and magnetization $m=1/3$; third, a flip of
the surface spins to the state $S_{i}=0$ with a modulated structure $%
\left\langle 3010\right\rangle $ (corresponding to the spin configuration $%
............\uparrow \uparrow \uparrow \bigcirc \uparrow \bigcirc \uparrow
\uparrow \uparrow \bigcirc \uparrow ....$) and magnetization $m=2/3$, and,
finally, the saturated state $\left\langle 3^{2}\right\rangle $ (corresponding
to the spin configuration $........\uparrow \uparrow \uparrow \uparrow \uparrow
\uparrow \uparrow \uparrow \uparrow .........$), with a magnetization $m=1$.


In order to obtain the values of the critical fields $h_{c_{1}}$ (transition
between the modulated structure $\left\langle 3\right\rangle $ and $%
\left\langle 3101\right\rangle $), $h_{c_{2}}$ (transition between the
modulated structure $\left\langle 3101\right\rangle $ and $\left\langle
3111\right\rangle $), $h_{c_{3}}$ (transition between the modulated
structure $\left\langle 3111\right\rangle $ and $\left\langle
3010\right\rangle $) and $h_{c_{s}}$ (transition between the modulated
structure $\left\langle 3010\right\rangle $ and $\left\langle
3^{2}\right\rangle $), we compare the energies for the respectiv periodic
sequence, finding the following critical fields:

\begin{equation}
\left\{ 
\begin{array}{c}
h_{c_1}=2+\delta \\ 
h_{c_2}=2-\delta \\ 
h_{c_3}=\alpha +\delta -1 \\ 
h_{c_s}=\alpha -\delta -1%
\end{array}
\right.  \label{27}
\end{equation}

By solving Eq.(26), we obtain the critical frontiers which separate the
various modulated phases, corresponding to magnetization plateaux at $m=1/6$%
, $1/3$ and $2/3$ (see discussion in the previous paragraph). We find that, for $0<h<h_{c1}$ and $%
h>h_{cs}$, the magnetizations are $m=0$ (disordered state) and $m=1$
(saturated state), respectively. Note that for $\delta <-2.0$ there is no
disordered state for positive field ($H>0$). Depending on the values of 
the parameters $\alpha $ and $\delta <0$, we can have various magnetization
plateaux at $m=1/6$, $1/3$ and $2/3$. In Figs 1, 2, and 3 the
ground state phase diagrams for $\delta =-1.0,-2.0$, and $-3.0$ are depicted,
respectively, where we indicate the various magnetization plateaux.

\section{Conclusions}

We have treated two soluble models: model $1$ (Eq. (6)) and 
model $2$ (Eq. (25)), by using a transfer-matrix technique. In the thermodynamical
limit ($N\rightarrow \infty $), the partition function was obtained and
ground state phase diagrams were calculated, in order to analyze the plateaux structure
in the magnetization. In contrast to the majority of existing approaches
(numerical diagonalization) to treat the problem of magnetization plateaux
in quantum models (see, for example, Ref.13), the present formalism
(transfer-matrix technique) is entirely based on analytical (exact)
calculations and allows for the calculation of magnetization profiles for arbitrary
finite temperatures and values of the parameters $\alpha $ and $\delta $.

\textbf{ACKNOWLEDGEMENT}

This work was partially supported by CNPq, FAPEAM\ and CAPES (Brazilian
Research Agencies).

\end{document}